\documentclass[aps,prb,twocolumn,a4paper,10pt,floatfix,superscriptaddress]{revtex4-1}
\pdfoutput=1

\usepackage{graphicx}
\usepackage{color}
\usepackage{textcomp}
\usepackage{amssymb}\usepackage{amsmath}\usepackage{amsthm}
\usepackage{units}
\usepackage[utf8]{inputenc}
\usepackage{times}

\begin{document}

\title{Short-range quantum magnetism\\ of ultracold fermions in an optical lattice}

\author{Daniel Greif}
\affiliation{Institute for Quantum Electronics, ETH Zurich, 8093 Zurich, Switzerland}

\author{Thomas Uehlinger}
\affiliation{Institute for Quantum Electronics, ETH Zurich, 8093 Zurich, Switzerland}

\author{Gregor Jotzu}
\affiliation{Institute for Quantum Electronics, ETH Zurich, 8093 Zurich, Switzerland}

\author{Leticia Tarruell}
\thanks{Present Address: ICFO-Institut de Ciencies Fotoniques, Mediterranean Technology Park, 08860 Castelldefels (Barcelona), Spain}
\affiliation{Institute for Quantum Electronics, ETH Zurich, 8093 Zurich, Switzerland}
\affiliation{LP2N UMR 5298, Univ. Bordeaux 1, Institut d’Optique and CNRS, 351 cours de la Libération, 33405 Talence, France}

\author{Tilman Esslinger}
\email{esslinger@phys.ethz.ch}
\affiliation{Institute for Quantum Electronics, ETH Zurich, 8093 Zurich, Switzerland}

\begin{abstract}
The exchange coupling between quantum mechanical spins lies at the origin of quantum magnetism. We report on the observation of nearest-neighbor magnetic spin correlations emerging in the many-body state of a thermalized Fermi gas in an optical lattice. The key to obtaining short-range magnetic order is a local redistribution of entropy within the lattice structure. This is achieved in a tunable-geometry optical lattice, which also enables the detection of the magnetic correlations. We load a low-temperature two-component Fermi gas with repulsive interactions into either a dimerized or an anisotropic simple cubic lattice. For both systems the correlations manifest as an excess number of singlets as compared to triplets consisting of two atoms with opposite spins. For the anisotropic lattice, we determine the transverse spin correlator from the singlet-triplet imbalance and observe antiferromagnetic correlations along one spatial axis. Our work paves the way for addressing open problems in quantum magnetism using ultracold fermions in optical lattices as quantum simulators.
\end{abstract}

\maketitle

Quantum magnetism describes quantum many-body states of spins coupled by exchange interactions and lies at the heart of many fundamental phenomena in condensed matter physics \cite{auerbach_interacting_1994,sachdev_quantum_2008}. 
While spin systems often tend to show long-range order at low temperatures, the fascinating interplay of exchange interactions with geometry and quantum fluctuations can lead to quantum states characterized by their short-range magnetic order. 
Examples include valence-bond crystals, spin-liquids and possibly high-temperature superconductors \cite{sachdev_quantum_2008,anderson_physics_2004,balents_spin_2010,auerbach_interacting_1994}. 
Quite remarkably, the underlying many-body physics gives rise to computationally and theoretically intractable regimes even in the phase diagrams of simple models, such as the Fermi-Hubbard model.
Moreover, the direct measurement of local spin correlations in solids remains a major challenge. 

The controlled setting of ultracold fermionic atoms in optical lattices is regarded as a promising route to gain new insights into phases driven by quantum magnetism \cite{lewenstein_ultracold_2007, bloch_many-body_2008, esslinger_fermi-hubbard_2010}. This approach offers experimental access to a clean and highly flexible Fermi-Hubbard model with a unique set of observables \cite{kohl_fermionic_2005}.
For repulsively interacting fermions density ordering in the metal-Mott insulator transition could already be explored experimentally \cite{jordens_mott_2008, schneider_metallic_2008}. 
Yet, progress towards entering the regime of quantum magnetism has been hindered by the ultra-low temperatures, and entropies, required to observe exchange-driven spin ordering in optical lattices. 
For bosonic quantum gases promising developments have been reported: by mapping the spin to other degrees of freedom the temperature limitation could be circumvented, which allowed the exploration of one-dimensional decoupled Ising spin chains and the simulation of classical magnetism on a triangular lattice \cite{simon_quantum_2011, struck_quantum_2011}.
Furthermore, exchange interactions were observed in artificially prepared arrays of isolated double-wells or plaquettes \cite{trotzky_time-resolved_2008, nascimbene_experimental_2012}. 

To overcome these limitations and directly access exchange-driven physics in thermalized systems, cooling schemes based on the redistribution of entropy between different regions of the trap have been suggested \cite{ho_squeezing_2009, bernier_cooling_2009}. In this work, we instead propose and implement a local entropy redistribution scheme within the lattice structure to reach the regime of quantum magnetism. The atoms are either prepared in a dimerized or an anisotropic simple cubic lattice, see Fig. 1. In both geometries, a subset of links of the underlying simple cubic lattice is set to a larger exchange energy as compared to the other links. As a result, the entropy is predominantly stored in configurations involving the weak links. For fixed total entropy in the trapped system, this essentially allows us to reach temperatures between the two exchange energy scales. In the dimerized lattice the resulting correlations on the strong links correspond to an excess population of the low energy singlet as compared to the triplet state -- in close resemblance to an explicit valence-bond crystal in the Heisenberg model \cite{diep_two-dimensional_2005}. In the anisotropic lattice the low temperatures lead to antiferromagnetic spin correlations along one spatial axis, the transverse component of which is also detected via a singlet-triplet imbalance. For both systems we study the dependence of the spin correlations as a function of temperature and tunneling balance and find good agreement with theory. 

\begin{figure}[ht]
\centering
\includegraphics[scale=0.9]{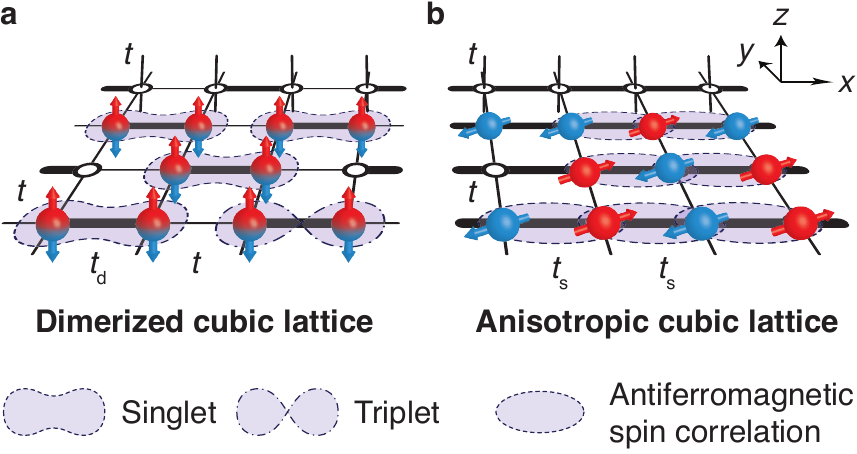}
\caption{
{\bf Magnetic spin correlations.} Schematic view of the nearest-neighbor spin correlations observed in the experiment. A two-component mixture of fermionic atoms (red and blue) is prepared close to half-filling in a cubic lattice with two different tunnel coupling configurations. {\bf a},~Dimerized lattice with the strong dimer links $t_{\mathrm{d}}$ and weaker links $t$. Low temperatures lead to an excess number of singlets over triplets. {\bf b},~Anisotropic lattice with strong and weak tunneling $t_{\mathrm{s}}$ and $t$ along different spatial axes. Antiferromagnetic spin correlations in the transverse direction are formed along the strong link direction. In both figures exemplary thermal excitations in the form of spin excitations or holes are shown.}
\label{fig:1}
\end{figure}

\begin{figure*}[ht]
\centering
\includegraphics[scale=0.9]{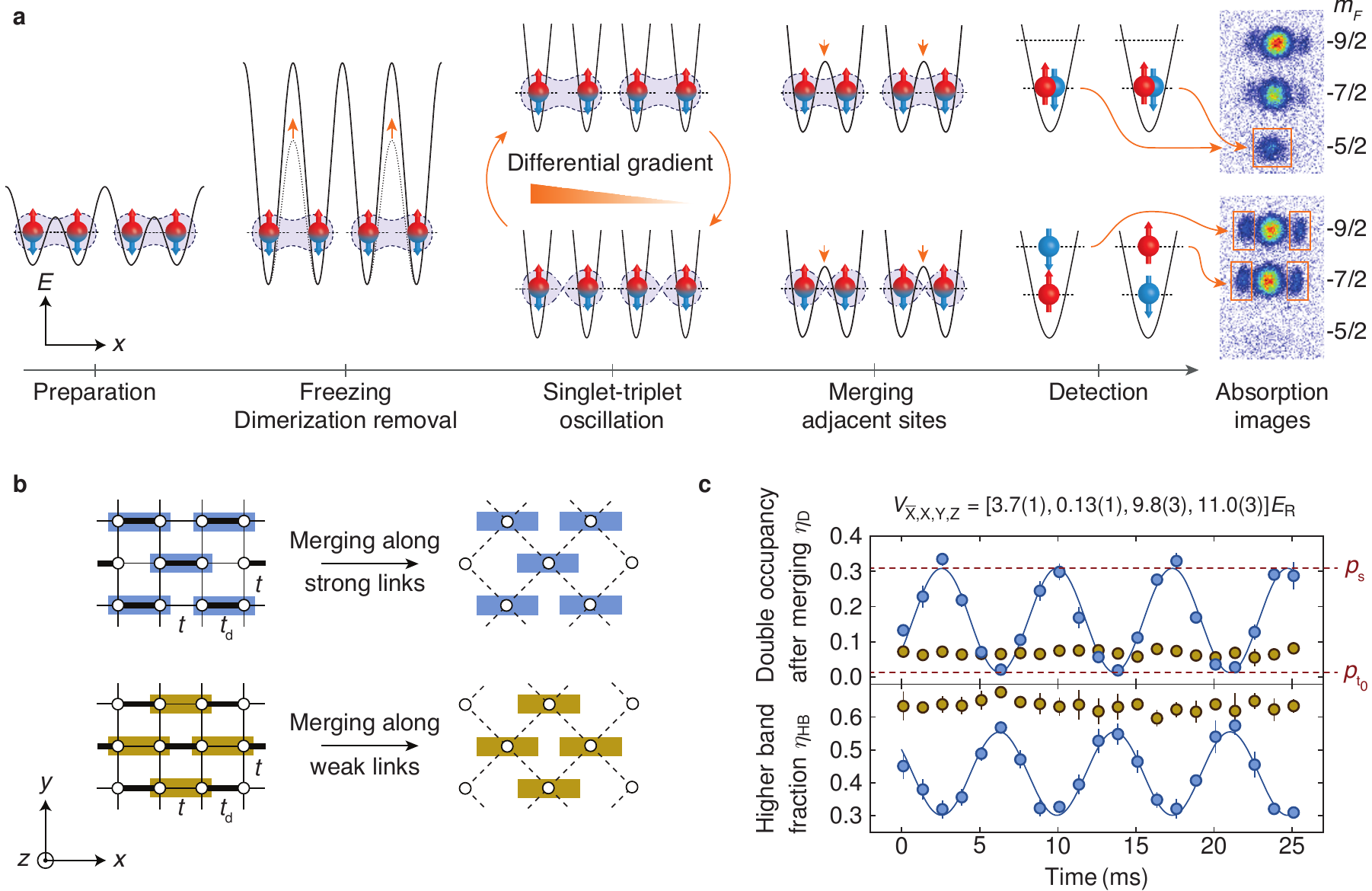}
\caption{
{\bf Detection scheme.} Summary of the technique used for measuring the atomic singlet and triplet fractions $p_{\mathrm{s}}$ and $p_{\mathrm{t_0}}$. {\bf a},~Schematic view of the different detection steps for the exemplary case of two singlet states in a dimerized lattice. Depending on the oscillation time, the absorption images on the right show either a large double occupancy in the lowest band corresponding to singlets (top row), or an increased higher band fraction indicating triplet states (bottom row). {\bf b},~The two possible merging configurations in a dimerized lattice. Singlets and triplets are detected on a set of adjacent sites arranged on a chequerboard pattern in the plane. {\bf c},~Exemplary singlet-triplet oscillation in a strongly dimerized lattice with $U/t=11.0(8)$ and $t_{\mathrm{d}}/t=22(2)$. We observe an oscillation in the double occupancy after merging, $\eta_{\mathrm{D}}$, and in the higher band fraction, $\eta_{\mathrm{HB}}$, when merging along the strong links (blue data), whereas no oscillations are visible for the weak links (ochre data). The phase of the oscillation is shifted owing to the double occupancy removal procedure \cite{supplementary}. The red dashed lines denote the extracted singlet and triplet fraction $p_{\mathrm{s}}$ and $p_{\mathrm{t_0}}$. Error bars show the standard deviation of at least five measurements.} \label{fig:2}
\end{figure*}

The experiment is performed with a harmonically confined, balanced two-component mixture of a quantum degenerate Fermi gas of $^{40}$K. The atoms are prepared in two magnetic sublevels $m_{F}=-9/2,-7/2$ of the $F=9/2$ hyperfine manifold, denoted by $\uparrow$ and $\downarrow$, at temperatures below $10\%$ of the Fermi temperature. We load $50,000-100,000$ repulsively interacting atoms at an $s$-wave scattering length of $106(1)\,a_0$ into the three-dimensional optical lattice, where $a_0$ denotes the Bohr radius. The lattice is created by a combination of interfering laser beams operating at $1064\,\mathrm{nm}$ with lattice depths $V_{\overline{\mathrm{X}}}$, $V_{\mathrm{X}}$, $V_{\mathrm{Y}}$ and $V_{\mathrm{Z}}$ \cite{tarruell_creating_2012, supplementary}. We independently control the tunneling strengths along all three spatial axes. In addition we can introduce a checkerboard dimerization in the $xy$ plane by strengthening every second tunneling link along the $x$ axis, see Fig. 1a. The checkerboard pattern replicates along the $z$ axis. Our system is well described by a three-dimensional single-band Hubbard model with repulsive on-site interaction energy $U$, unless explicitly stated. The tunneling along the weak links in both lattice geometries is set to $t/h=67(3)\,\mathrm{Hz}$ for all measurements, where $h$ denotes Planck's constant.

As shown in Fig.~2a and b, the fraction of atoms forming singlets and triplets on neighboring lattice sites ($p_{\mathrm{s}}$ and $p_{\mathrm{t_0}}$) is detected by transforming the lattice to a checkerboard geometry, similar to a previously developed technique \cite{trotzky_controlling_2010}. In the dimerized lattice our detection scheme locally projects onto the two-site eigenstates of isolated dimers, whereas for the anisotropic lattice the singlet state is directly projected onto $(\left|\uparrow,\downarrow\right\rangle-\left|\downarrow,\uparrow\right\rangle)/\sqrt{2}$ (Ref.~\onlinecite{supplementary}). In the first detection step, the atomic motion in the initial lattice is frozen by rapid conversion to a simple cubic structure with negligible tunneling. Next, all atoms on doubly occupied sites are removed. We then apply a magnetic field gradient, which creates a differential bias energy $\Delta$ for atoms of opposite spins on adjacent sites and causes coherent oscillations between the singlet $\left|\mathrm{s}\right\rangle=(\left|\uparrow,\downarrow\right\rangle-\left|\downarrow,\uparrow\right\rangle)/\sqrt{2}$ and the triplet $\left|\mathrm{t_0}\right\rangle=(\left|\uparrow,\downarrow\right\rangle+\left|\downarrow,\uparrow\right\rangle)/\sqrt{2}$ state at a frequency $\nu = \Delta /h$. If the initial amount of singlets and triplets is equal, no overall oscillation will be visible, as $\left|\mathrm{s}\right\rangle$ and $\left|\mathrm{t_0}\right\rangle$ oscillate in antiphase.

After a certain oscillation time, we remove the gradient and merge two adjacent sites. Owing to the symmetry of the two-particle wavefunction, the singlet state on neighboring sites evolves to a doubly occupied site with both atoms in the lowest band, while the triplet state transforms into a state with one atom in the lowest and one atom in the first excited band. The fraction of atoms forming double occupancies in the lowest band of the merged lattice,$\eta_{\mathrm{D}}$, is detected by a radiofrequency transfer to the previously unpopulated $m_{F}=-5/2$ spin state \cite{jordens_mott_2008}. The fraction of atoms in the higher band $\eta_{\mathrm{HB}}$ is obtained from a band mapping technique \cite{esslinger_fermi-hubbard_2010}. For the final readout we take absorption images after Stern-Gerlach separation of the spin states during ballistic expansion. For an imbalance between the initial singlet and triplet populations, $\eta_{\mathrm{D}}$ and $\eta_{\mathrm{HB}}$ will show oscillations with opposite amplitudes. As the double occupancy in the lowest band contains only contributions from two particles with opposite spins, we can infer the fraction of atoms forming singlets and triplets from the maxima and minima of a sinusoidal fit to $\eta_{\mathrm{D}}$. The higher band fraction has an additional offset caused by dimers containing two atoms with the same spin or one atom with an antisymmetric spatial wavefunction.

\begin{figure}[ht]
\centering
\includegraphics[scale=0.9]{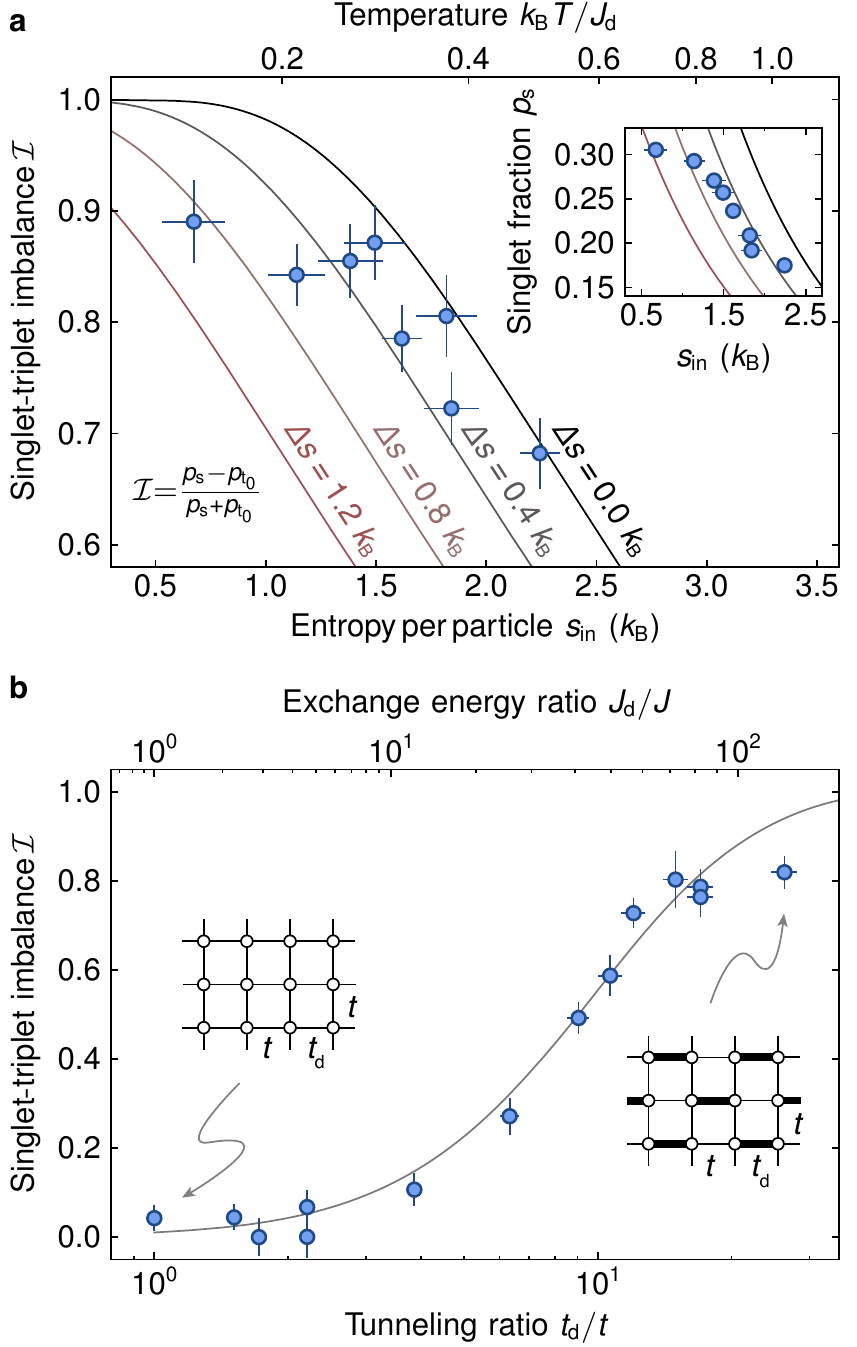}
\caption{
{\bf Dimerized simple cubic lattice.} 
{\bf a},~Singlet-triplet imbalance on the strong dimer links vs. initial entropy before loading into the lattice $s_{\mathrm{in}}$ and temperature $k_{\mathrm{B}}T/J_{\mathrm{d}}$ in a dimerized lattice with $U/t=11.0(8)$ and $t_{\mathrm{d}}/t=22(2)$. The imbalance and the absolute singlet fraction (inset) decrease with increasing entropy. Solid lines are the prediction of a high-temperature series expansion taking into account different amounts of added entropy $\Delta s$ during the lattice loading. {\bf b},~Imbalance versus dimerization $t_{\mathrm{d}}/t$ and $J_{\mathrm{d}}/J$, showing an increase for strongly dimerized simple cubic lattices. The solid line is the theory prediction for an entropy per particle of $1.8\,k_{\mathrm{B}}$ in the lattice, which includes the heating during loading. Vertical error bars denote the fit error from singlet-triplet oscillations consisting of 63 measurements, the errors in $t_{\mathrm{d}}/t$ stem from lattice calibration uncertainties and the errors in $s_{\mathrm{in}}$ are the standard deviation of five measurements. For individual curves of $p_{\mathrm{s}}$ and $p_{\mathrm{t_0}}$ for all measurements see \cite{supplementary}.} \label{fig:3}
\end{figure}

When loading atoms into a strongly dimerized lattice and merging along the strong links, we observe oscillations in $\eta_{\mathrm{D}}$ and $\eta_{\mathrm{HB}}$, see Fig. 2c. This reveals an excess number of singlets, corresponding to magnetic order on neighboring sites. We quantify this order by the normalized imbalance
\begin{equation}
\mathrm{\cal{I}}=\frac{p_{\mathrm{s}}-p_{\mathrm{t_0}}}{p_{\mathrm{s}}+p_{\mathrm{t_0}}}.
\end{equation}
The order in the strongly dimerized lattice originates from temperatures below the intra-dimer exchange energy $J_{\mathrm{d}}=-U/2+\sqrt{16t_{\mathrm{d}}^2+U^2}/2$, which denotes the singlet-triplet splitting on a single dimer. While such temperatures require very low entropies for isotropic lattices \cite{fuchs_thermodynamics_2011}, in our system the access to the regime of magnetic ordering is facilitated by the presence of the weaker inter-dimer exchange energy $J\ll J_{\mathrm{d}}$. This leads to an entropy redistribution from states on the strong to the weak links and gives access to the temperature regime $J<k_{\mathrm{B}}T<J_{\mathrm{d}}$ for experimentally attainable entropies (here $k_{\mathrm{B}}$ denotes the Boltzmann constant). As expected for strong dimerization, we find no visible oscillations when merging along the weak links, which indicates the absence of magnetic correlations on these links, see Fig. 2c. The observed constant values of $\eta_{\mathrm{D}}=0.07(1)$ and $\eta_{\mathrm{HB}}=0.63(3)$ are consistent with a state where nearly all singlets are located on neighboring dimer links, with vanishing correlations between them. 

To analyze the effect of temperature on the magnetic correlations, we measure the dependence of the singlet-triplet imbalance on entropy, see Fig. 3a. The imbalance $\mathrm{\cal{I}}$ and the absolute singlet fraction $p_{\mathrm{s}}$ reduce for larger entropies, as triplet states become thermally populated. The singlet fraction is additionally diminished by a shrinking half-filled region in the trapped system~\cite{greif_probing_2011}. We find good agreement with a second order high-temperature series expansion of coupled dimers when including an entropy increase of $\Delta s=0.4\,k_{\mathrm{B}}$ with respect to the initial entropy in the harmonic trap, $s_{\mathrm{in}}$. This heating is associated to the lattice loading \cite{jordens_quantitative_2010} and is larger for the lowest entropies, consistent with previous results \cite{greif_probing_2011}. From the measured imbalances we infer temperatures below $0.4J_{\mathrm{d}}$.

For reduced dimerizations the coupling between dimers leads to increased inter-dimer correlations. The excitation energy of triplets is then lowered as they delocalize over the lattice, thus changing the nature of magnetic ordering. In Fig. 3b we use the tunable lattice to investigate the dependence of the imbalance $\mathrm{\cal{I}}$ on the tunneling ratio $t_\mathrm{d}/t$. As the dimerization is progressively removed the imbalance decreases in good agreement with theory and eventually falls below our experimental resolution. This decrease can be attributed to the inter-dimer exchange energy $J_{\mathrm{d}}$ becoming smaller than the temperature $T$. For vanishing temperatures the system is expected to undergo a quantum phase transition from a gapped spin-liquid state to a long-range ordered antiferromagnet as $t_\mathrm{d}/t$ is reduced below a critical value, where the spin gap becomes zero \cite{sachdev_quantum_2008, ruegg_pressure-induced_2004}. 

\begin{figure}[ht]
\centering
\includegraphics[scale=0.9]{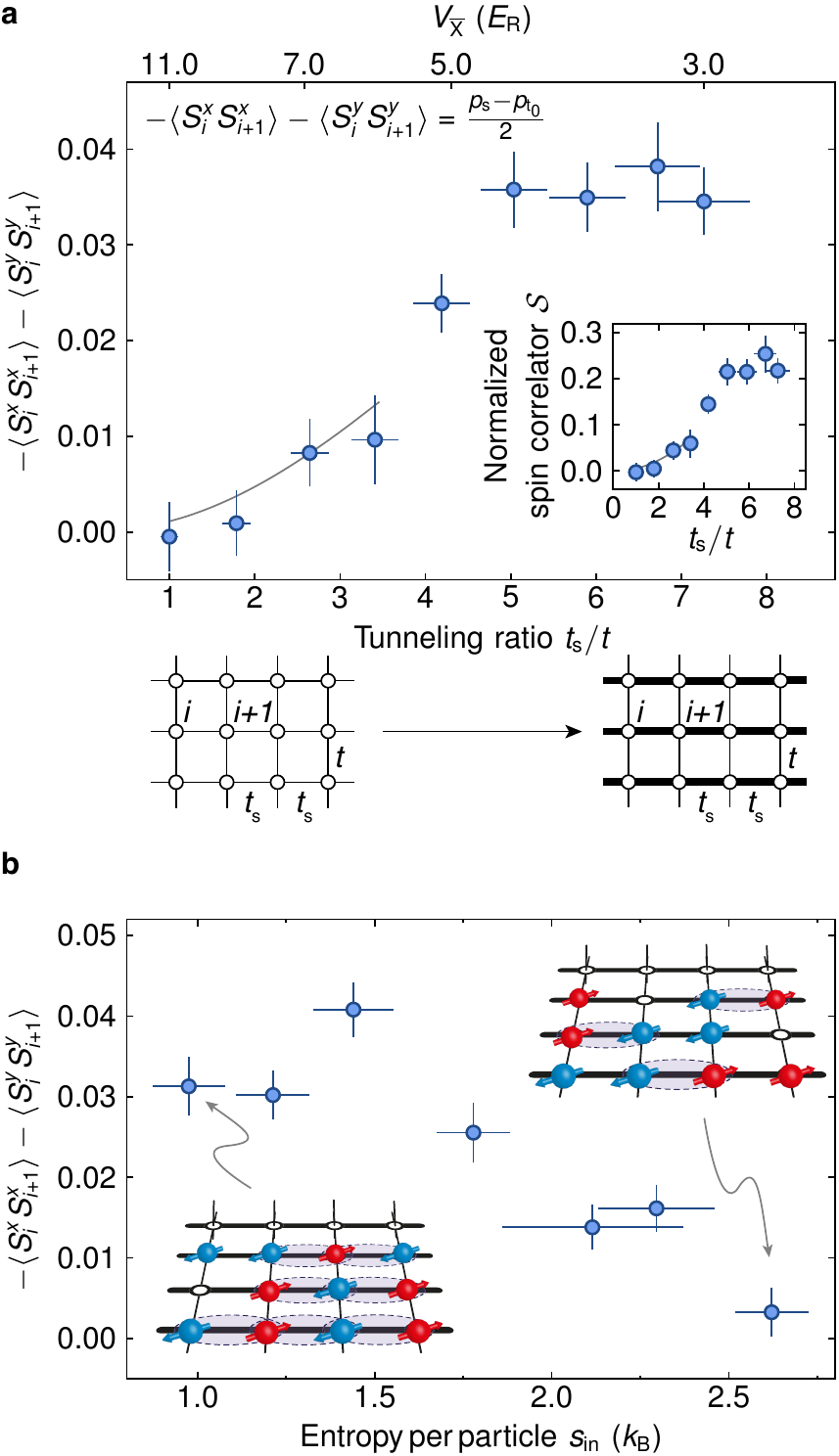}
\caption{
{\bf Nearest-neighbor antiferromagnetic order.} 
{\bf a},~Transverse spin correlator versus tunneling ratio $t_s/t$ and lattice depth $V_{\overline{\mathrm{X}}}$ in a three-dimensional anisotropic simple cubic lattice with $V_{\mathrm{Y},\mathrm{Z}}=11.0(3)\,E_{\mathrm{R}}$. Positive values correspond to antiferromagnetic ordering. The inset shows the normalized spin correlator~$\mathrm{\cal{S}}$, denoting the fraction of antiferromagnetic ordering at the relevant density. Here $U/t$ decreases from $16(1)$ to $10.5(8)$. Solid lines are the prediction of a high-temperature series expansion for an entropy per particle of $1.8\,k_{\mathrm{B}}$, as used in Fig.~3b, and are shown up to $t_{\mathrm{s}}/k_{\mathrm{B}}T=1/2$. {\bf b},~Transverse spin correlator versus entropy before loading into the lattice at $U/t=10.5(8)$ and $t_{\mathrm{s}}/t=7.3(6)$, together with a schematic view of the spin ordering. Error bars as in Fig.~3.}
\label{fig:4}
\end{figure}

The key to the observation of quantum magnetism in our system is the presence of two different exchange energy scales. Without dimerization, this situation also occurs for anisotropic simple cubic lattices with tunneling~$t$ along two axes and a stronger tunneling $t_{\mathrm{s}}$ along the third direction. In this case the symmetry between neighboring links is restored and the detected singlet and triplet fractions are the same for both merging configurations. We observe a clear population difference $(p_{\mathrm{s}}-p_{\mathrm{t_0}})/2$ after loading a gas with entropies $s_{\mathrm{in}}$ below $1.0\,k_{\mathrm{B}}$ into an anisotropic lattice, which increases to $4\%$ for larger tunneling ratios $t_{\mathrm{s}}/t$, see Fig. 4a. The population difference is equal to the transverse spin correlator between neighboring sites $i$ and $i+1$ along the strong tunneling direction
\begin{equation}
-\langle S^{x}_{i}S^{x}_{i+1}\rangle-\langle S^{y}_{i}S^{y}_{i+1}\rangle=\left(p_{\mathrm{s}}-p_{\mathrm{t_0}}\right)/2.
\label{eq:PsPtSxSy}
\end{equation}
This quantity hence directly characterizes the fraction of atoms with antiferromagnetic ordering on neighboring sites in the entire atomic cloud. Our observations also extend to weak lattices, where correction terms to the single-band Hubbard model become relevant \cite{werner_interaction-induced_2005}. In this regime a variety of magnetic phases have been predicted \cite{mathy_accessing_2009, ma_density_2012}. 

The results can be compared to a second order high-temperature series expansion \cite{supplementary}. We find good agreement in the regime of small anisotropies. For larger anisotropies the expansion breaks down, as the strong tunneling and the temperature become comparable. In this regime we expect the temperature to lie between the large and small exchange scales $J<k_{\mathrm{B}}T<J_{\mathrm{s}}$. The system then behaves as an array of one-dimensional spin-ordered chains without correlations between them \cite{giamarchi_quantum_2003}, where the majority of the entropy is stored in configurations involving the weak links. Low-dimensional systems have been predicted to show enhanced nearest-neighbour correlations \cite{gorelik_universal_2012}.

For temperatures much larger than the strong exchange energy the magnetic correlations should disappear. In Fig. 4b we study the dependence on the initial entropy $s_{\mathrm{in}}$ and find the correlations to vanish above $2.5\,k_{\mathrm{B}}$, where $k_{\mathrm{B}}T\gg J_{\mathrm{s}}$ is expected. 

Owing to the presence of the harmonic trap, most spin correlated atoms are located in the center, where the filling is close to one particle per site. The density-normalized fraction of antiferromagnetic ordering is obtained when dividing by the fraction of atoms with two particles of arbitrary spin on adjacent sites. Under the assumption that all spin correlators $\langle S^{x,y,z}_{i}S^{x,y,z}_{i+1}\rangle$ are equal -- which applies if all symmetry breaking fields are much smaller than all other energy scales -- the normalized spin correlator $\cal{S}$ can be directly obtained from the measurement of singlets and triplets (here $n^{\mathrm{s}}_{i}$ is one for a single particle of any spin on site $i$ and zero otherwise)
\begin{equation}
\mathrm{\cal{S}}=\frac{-4\langle S^{z}_{i}S^{z}_{i+1}\rangle}{\langle n^{\mathrm{s}}_{i}n^{\mathrm{s}}_{i+1}\rangle}=\frac{p_{\mathrm{s}}-p_{\mathrm{t_0}}}{p_{\mathrm{s}}+3p_{\mathrm{t_0}}}. 
\end{equation}
The normalized antiferromagnetic correlations along the strong tunneling direction reach $25\%$, see inset Fig.~4a. This corresponds to approximately $5,000$ ordered atoms. 

In this work, we have demonstrated the observation of short-range quantum magnetism of repulsively interacting ultracold fermions in cubic lattices and investigated the dependence on temperature, lattice dimerization and anisotropy. Our approach is based on a local entropy redistribution scheme within the lattice structure and can be generalized to access the low temperature regime in different geometries, for example two-dimensional systems. The tunable geometry optical lattice allows the extension of our studies to spin-ladder systems, dimerized one-dimensional chains and zig-zag chains, where the interplay between quantum fluctuations and magnetic ordering plays a particularly important role \cite{giamarchi_quantum_2003, he_magnetic_2007}. At even lower temperatures, the existence of spin-liquids in honeycomb or triangular lattices could be investigated \cite{meng_quantum_2010}.

\textbf{Acknowledgments} We would like to thank Nils Blümer, Thierry Giamarchi, Corinna Kollath, Henning Moritz, Christian Rüegg, Manfred Sigrist, Niels Strohmaier and Shun Uchino for insightful discussions and Ulf Bissbort for help with the calculation of the Hubbard parameters in the dimerized lattice. We acknowledge SNF, NCCR-QSIT, NCCR-MaNEP, and SQMS (ERC advanced grant) for funding.

\clearpage

\section*{Supplementary Materials}
\setcounter{section}{0}
\setcounter{subsection}{0}
\setcounter{figure}{0}
\setcounter{equation}{0}
\renewcommand{\figurename}[1]{FIG. S}
\renewcommand{\theequation}{S\arabic{equation}}

\section{Experimental Details}

\subsection{Preparation}
After sympathetic cooling with $^{87}$Rb, $2\times10^6$ fermionic $^{40}$K atoms are transferred into an optical dipole trap operating at a wavelength of $826\,\text{nm}$. A balanced incoherent spin mixture of atoms in the Zeeman levels $m_F=-9/2$ and $-7/2$ of the $F = 9/2$ hyperfine manifold is then prepared and evaporatively cooled \cite{jordens_mott_2008}. When taking data as a function of entropy, the gas is heated through inelastic losses by setting the magnetic bias field to a value close to the Feshbach resonance at $202.1\,\mathrm{G}$. We measure the entropy per particle in the dipole trap $s_{\mathrm{in}}$ using Fermi fits to the momentum distribution of the cloud after expansion. The field is finally increased to $221.4\,\mathrm{G}$ resulting in a scattering length of $106(1)\,a_0$. The optical lattice is subsequently turned on in $200\,\mathrm{ms}$ using a spline shaped laser-intensity ramp.

\subsection{Trapping potential}

The red-detuned optical lattice is created by four retro-reflected laser beams at $\lambda=1064\,\mathrm{nm}$ \cite{tarruell_creating_2012}. This gives rise to a potential of the form 
\begin{eqnarray} 
V(x,y,z) & = & -V_{\overline{\mathrm{X}}}\cos^2(kx+\theta/2)-V_{\mathrm{X}} \cos^2(kx)\nonumber\\
&&-V_{\mathrm{Y}} \cos^2(k y)\nonumber\\
&&-2\alpha \sqrt{V_{\mathrm{X}}V_{\mathrm{Y}}}\cos(kx)\cos(ky)\cos\varphi \nonumber\\
&&-V_{\mathrm{Z}} \cos^2(kz), \label{eqlattice}
\end{eqnarray}
where $V_{\overline{\mathrm{X}}}$, $V_{\mathrm{X}}$, $V_{\mathrm{Y}}$ and $V_{\mathrm{Z}}$ denote single-beam lattice depths (as calibrated using Raman-Nath diffraction on a $^{87}$Rb Bose-Einstein condensate), $k=2\pi/\lambda$ and the measured visibility of the interference pattern $\alpha$ is $0.90(5)$. The phase $\varphi$ is stabilized to $0.00(3)\pi$, whilst $\theta$ is set to $1.000(1)\pi$. Gravitation points along the $y$~direction.

The laser beams create an overall harmonic trapping potential which scales with the lattice depths according to the approximate expressions 
\begin{eqnarray}
\omega_x&\propto&\sqrt{V_{\mathrm{Y}}+1.11V_{\mathrm{Z}}} \nonumber \\
\omega_y&\propto&\sqrt{V_{\overline{\mathrm{X}}}+0.81(V_{\mathrm{X}}V_{\mathrm{Y}}/V_{\overline{\mathrm{X}}})+0.78V_{\mathrm{Z}}} \nonumber \\
\omega_z&\propto&\sqrt{V_{\overline{\mathrm{X}}}+0.81(V_{\mathrm{X}}V_{\mathrm{Y}}/V_{\overline{\mathrm{X}}})+1.24V_{\mathrm{Y}}}.
\end{eqnarray}
For $V_{{\overline{\mathrm{X}}},\mathrm{X},\mathrm{Y},\mathrm{Z}}=[3.7(1),0.13(1),9.8(3),11.0(3)]\,E_{\mathrm{R}}$, as in Fig. 3a, the lattice contributes trapping frequencies of $\omega^{\mathrm{Lattice}}_{x,y,z}/2\pi=[62.7(9), 57(1),54.3(3)]\,\mathrm{Hz}$. Additionally, the optical dipole trap creates a trapping of $\omega^{\mathrm{Dipole}}_{x,y,z}/2\pi=[30.7(2),105.9(3),34.6(2)]\,\mathrm{Hz}$, in all data sets shown, except for Fig. 2c and 4a, where $\omega^{\mathrm{Dipole}}_{x,y,z}/2\pi=[28.1(2),90.1(3),31.6(2)]\,\mathrm{Hz}$.

\subsection{Detection lattice ramp}
For measurements in the dimerized lattice, the lattice is ramped up in two steps. All beam intensities are linearly increased over the course of $0.5\,\mathrm{ms}$ up to the point where $V_{\mathrm{Y}}= 30(1)\,E_{\mathrm{R}}$, $V_{\mathrm{Z}} = 40(1)\,E_{\mathrm{R}}$ and all other intensities in the $xy$ plane are ramped such that the potential is not deformed. In a second linear ramp lasting $10\,\mathrm{ms}$, the lattice is changed to a simple cubic geometry where $V_{{\overline{X}},X,Y,Z}=[25(1),0,30(1),40(1)]\,E_{\mathrm{R}}$.

This ramp can be considered sudden for the inter-dimer links but adiabatic for the intra-dimer links. Our observable hence locally projects onto the two-site eigenstates of individual dimers, which includes an admixture of double occupancies. We use an exact calculation of a two-site Hubbard model to estimate the adiabaticity of the ramp. The unnormalized singlet ground state of this system is given by 
\begin{eqnarray} \frac{4t_{\mathrm{d}}}{-U + \sqrt{16 t_{\mathrm{d}}^2 + U^2} } \left(\left|\uparrow,\downarrow\right>-\left|\downarrow,\uparrow\right>\right) + \left(\left|\uparrow\downarrow,0\right> + \left|0,\uparrow\downarrow\right>\right). \nonumber \\
\label{fullsinglet}
\end{eqnarray}
There is hence a significant contribution of double occupancy in the regime where $U\sim t_{\mathrm{d}}$, as applies for the most strongly dimerized lattices investigated in this paper, whilst the contribution vanishes for the deep simple cubic lattice used for detection where $U/t\approx600$. For the given ramp-times and including site-offsets due to the harmonic trapping potential, the probability of populating excited states on a dimer during such a ramp remains below $0.1\%$ for all values of $U$ and $t_{\mathrm{d}}$ explored in the dimerized systems. 

In the anisotropic lattice, we directly ramp to $V_{{\overline{\mathrm{X}}},\mathrm{Y},\mathrm{Z}}=[25(1),30(1),40(1)]\,E_{\mathrm{R}}$ in $0.5\,\mathrm{ms}$ ($V_{\mathrm{X}} = 0\,E_{\mathrm{R}}$ throughout). In contrast to the dimer lattice ramp, this process can be considered sudden for all links, as the duration of the ramp is always well below the tunneling time in the initial lattice. Our detection method then corresponds to locally projecting the wavefunction of the system onto $(\left|\uparrow,\downarrow\right>-\left|\downarrow,\uparrow\right>)/\sqrt{2}$ on pairs of sites when measuring $p_{\mathrm{s}}$, hence excluding any contributions from double occupancies. The probability of this projection, as calculated from the two-site Hubbard model, lies above $80\%$ for all shown data and is higher for deeper lattices. For both lattice geometries, the triplet state remains unaffected by changes in $U$ and $t$.

\subsection{Singlet-triplet oscillations}

For both systems, once the ramp to the deep simple cubic lattice has been completed, double occupancies are removed via spin-changing collisions, which occur after transferring atoms from the $m_F=-7/2$ to the $m_F=-3/2$ state\cite{krauser_coherent_2012}. We verify that this procedure removes all double occupancies but leaves singly occupied sites unaffected by measuring the number of double occupancies and the total number of atoms before and after applying the removal sequence. A magnetic field gradient causing spin-dependent energy offsets of $\Delta_{-9/2,-7/2}/h=[1291(1),1156(1)]\,\mathrm{Hz}$ between neighboring sites is then applied, giving rise to coherent oscillations between singlets and triplets \cite{trotzky_controlling_2010}. Subsequently, pairs of adjacent lattice sites are adiabatically merged into one by linearly reducing $V_{\overline{\mathrm{X}}}$ to zero whilst increasing $V_{\mathrm{X}}$ to $25(1)\,E_{\mathrm{R}}$. The double occupancy in the merged lattice is then measured as \cite{jordens_mott_2008}, but taking into account an independently calibrated detection efficiency of $89(2)\%$. We verify that merged sites containing two atoms of opposite spin but in different bands are not detected as double occupancies by artificially creating a state containing large amounts of triplets but no singlets and measuring $\eta_\mathrm{D}$.

We apply a sinusoidal fit to the double occupancy where the frequency and phase are fixed and take into account the damping of the oscillations, which was calibrated independently and is included by multiplying the amplitude by $1.16$. A phase shift arises owing to a weak residual magnetic field gradient present during the double occupancy removal procedure, whereas the contribution from switching the singlet-triplet oscillation gradient on and off is negligible. We confirm that the maximum of the oscillation corresponds to its starting point (and hence to the number of singlets) by merging the lattice immediately after it has been ramped to a deep simple cubic structure.

\subsection{Theoretical model}
For sufficiently deep lattices the system is well described by a single band Fermi-Hubbard model, which is given by
\begin{eqnarray}
\hat H=
    -t_{i,j}\sum_{\langle ij\rangle,\sigma}
    (\hat c^\dagger_{i\sigma}\hat c_{j\sigma}+\text{h.c.})
    +U\sum_i \hat n_{i\uparrow}\hat n_{i\downarrow}
\label{eq:fermi-hubbard-general}
\end{eqnarray}
for a homogeneous system, using standard notation.The interaction energy $U$ and nearest-neighbor tunneling $t_{i,j}$ are evaluated from Wannier function integrals \cite{lewenstein_ultracold_2007}. The dimerized lattice theory curves are calculated from a perturbative coupling of isolated dimer links with tunneling $t$. The resulting partition function expansion is then calculated up to second order in $t/k_{\mathrm{B}}T$. While the dimer link contains 16 states in total and the tunneling operator is non-diagonal between neighboring dimers, the evaluation of the relevant matrix elements is directly analogous to the case of single sites \cite{henderson_high-temperature_1992}. The harmonic trap is included in a local density approximation, which leads to a quadratically varying chemical potential. All thermodynamic quantities are obtained after integration over the entire trap using independent calibrations of the atom number, trap frequencies and lattice depths \cite{jordens_quantitative_2010}. In the limit of very strong dimerization, theory predicts $p_{\mathrm{t_0}}/p_{\mathrm{s}} \propto \mathrm{exp}(-J_{\mathrm{d}}/k_{\mathrm{B}}T)$, which can be used for lattice thermometry.

For the anisotropic lattice we evaluate the correlators $\langle S^{\mathrm{z}}_{\mathrm{i}}S^{\mathrm{z}}_{\mathrm{i+1}}\rangle$ and $\mathrm{\cal{S}}$ in a second order series expansion of coupled single sites \cite{ten_haaf_high-temperature_1992}. The thermodynamic observables are obtained in a similar way as previously described \cite{jordens_quantitative_2010}, using the average tunneling $\sqrt{(t_{\mathrm{s}}^2+2t^2)/3}$. Eq.~(2) in the main text is computed by evaluating the matrix elements of the spin operators.

\section{Dimerized cubic lattice - series expansion}

We start with the definition of the homogeneous single-band Hubbard Hamiltonian in a dimerized cubic lattice with strong and weak links between nearest neighbors $\left<i,j\right>_{\linethickness{2pt}\line(1,0){5}}$ and $\left<i,j\right>_{\line(1,0){5}}$ respectively, see Fig. S1,
\begin{eqnarray}
\hat H_{D}&=&\hat H_{o}+\hat H_{c}\nonumber \\
    \hat H_{o}&=&-t_d\sum_{\sigma,\left<i,j\right>_{\linethickness{2pt}\line(1,0){5}}} 
    (\hat c^\dagger_{i,\sigma}\hat c_{j,\sigma}+\text{h.c.}) \nonumber \\ & &
    +U\sum_{i} \hat n_{i\uparrow}\hat n_{i\downarrow}-\mu\sum_{i}
    ( \hat n_{i\uparrow}+\hat n_{i\downarrow}) \\
    \hat H_{c}&=&-t\sum_{\sigma,\left<i,j\right>_{\line(1,0){5}}}
    (\hat c^\dagger_{i,\sigma}\hat c_{j,\sigma}+\text{h.c.}).\nonumber
\label{eq:fermi-hubbard}
\end{eqnarray}
We have split the Hamiltonian into the dimer part $\hat H_o$ and the coupling between dimers $\hat H_c$. The on-site interaction energy is given by $U$, the tunneling matrix elements between nearest neighbors by $t$ and $t_d$ and the chemical potential is parametrized with $\mu$. The fermionic creation operator for an atom on the lattice site $i$ is given by $\hat c^\dagger_{i\sigma}$, where $\sigma\in\{\uparrow,\downarrow\}$ denotes the magnetic sublevel and h.c. is the Hermitian conjugate. The particle number operator is $\hat n_i=\hat n_{i\uparrow}+\hat n_{i\downarrow}$, $\hat n_{i\sigma}=\hat c^\dagger_{i\sigma}\hat c_{i\sigma}$. Denoting the inverse temperature with $\beta=1/k_{\mathrm{B}}T$, the thermal average of an observable $\mathrm{\cal{O}}$ then reads in the grand canonical potential

\begin{figure}[ht]
\includegraphics[scale=0.9]{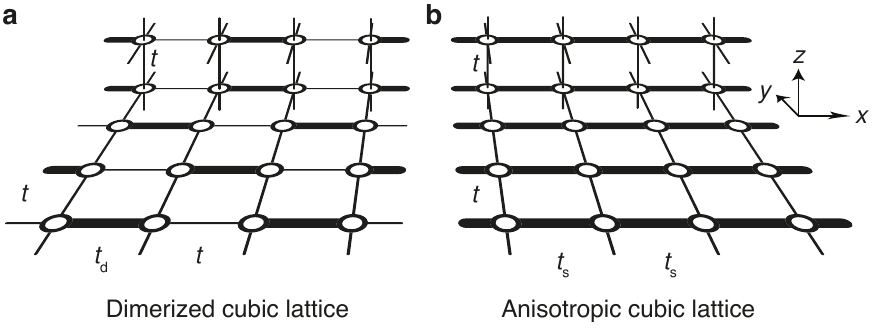}
\caption{
{\bf Lattice geometries.} Overview of the three-dimensional lattice geometries explored in the experiment. {\bf a,}~In the dimerized cubic lattice the tunneling between neighboring sites is increased to $t_{\mathrm{d}}$ on a checkerboard pattern in the $xy$ plane as compared to the weaker tunneling $t$. The pattern replicates along the $z$ axis. {\bf b,}~In the anisotropic cubic lattice the tunneling is increased to $t_{\mathrm{s}}$ along the $x$ axis, while remaining $t$ in the other two directions.} 
\label{sfig:1}
\end{figure}

\begin{equation}
\label{eq:obs}
\langle \mathrm{\cal{O}} \rangle = \frac{\mbox{Tr}\{\mathrm{\cal{O}}e^{-\beta\hat H_{D}}\}}{\mbox{Tr}\{e^{-\beta\hat H_{D}}\}}.
\end{equation}
We now treat the coupling Hamiltonian $\hat H_c=t\hat T$ as a perturbation, which leads to an expansion of the above expression in powers of the dimensionless parameter $\beta t$ \cite{henderson_high-temperature_1992}. The expansion is expected to be close to the exact result in the regime $t\ll k_{\mathrm{B}}T\ll t_d, U$. For the partition function up to second order (denominator in Eq. \ref{eq:obs}) we find
\begin{equation}
\label{eq:hts}
\mathrm{\cal{Z}}=\mathrm{\cal{Z}}_0+(\beta t)^2\frac{\mathrm{\cal{Z}}_0}{\beta^2}\int_0^\beta\int_0^{\tau_1}\mathrm{d}\tau_2\mathrm{d}\tau_1\langle\hat T'(\tau_1)\hat T'(\tau_2)\rangle_0.
\end{equation}
The expression for the numerator is analogous. The partition function of the unperturbed Hamiltonian is denoted by $\mathrm{\cal{Z}}_0$, whereas $\langle...\rangle_0$ denotes the thermal average of the unperturbed Hamiltonian
\begin{eqnarray}
\langle\hat T'(\tau_1)\hat T'(\tau_2)\rangle_0&=&\mathrm{Tr}\{\mathrm{exp}(-\beta\hat H_{o})\hat T'(\tau_1)\hat T'(\tau_2)\}/\mathrm{\cal{Z}}_0 \nonumber \\
\hat T'(\tau)&=&e^{\tau\hat H_o} \hat T e^{-\tau\hat H_o}.
\end{eqnarray}

\section{Dimerized cubic lattice - observables}

\begin{figure*}[ht!]
\includegraphics[scale=0.75]{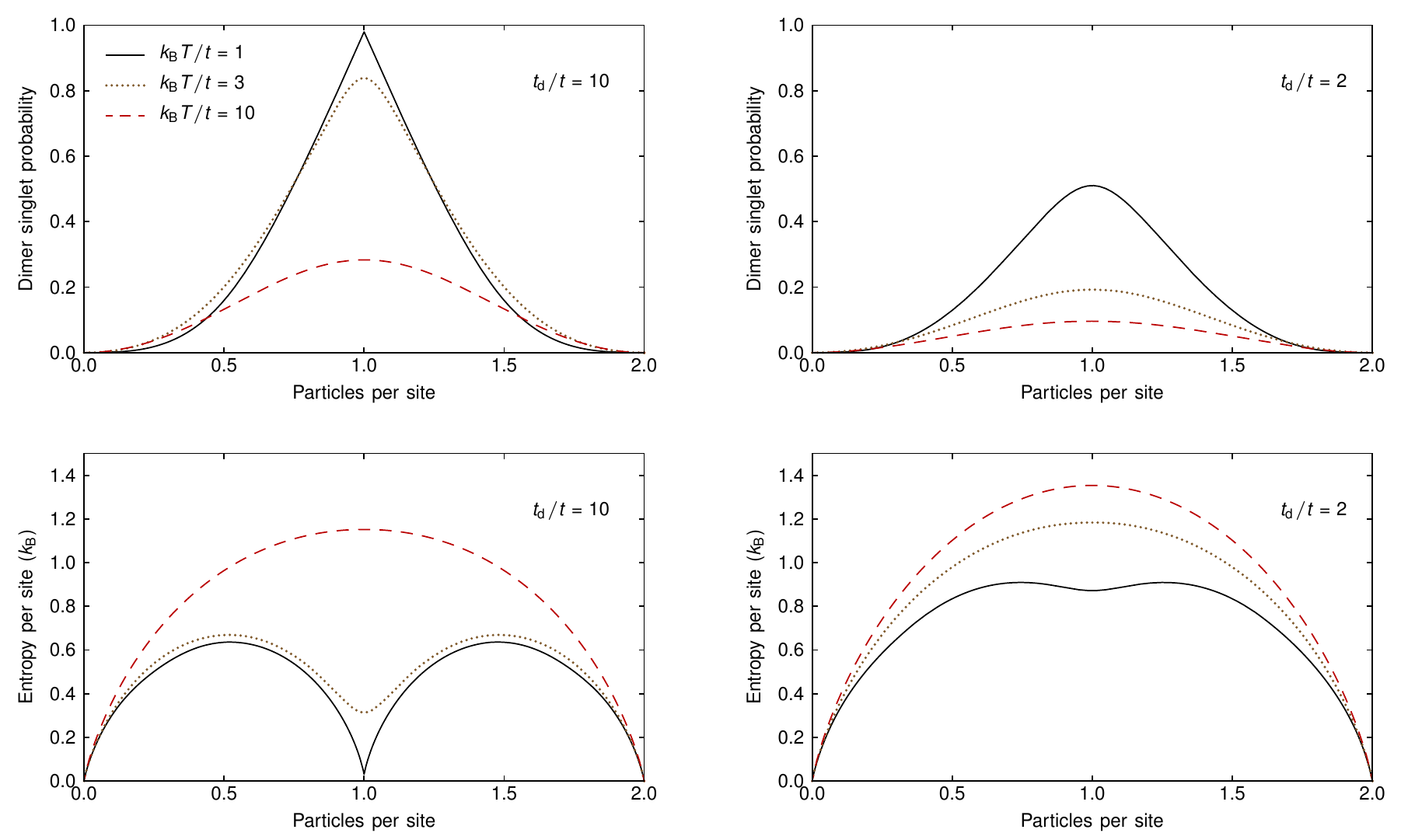}
\caption{
{\bf Dimerized lattice.} High-temperature series predictions up to second order for the homogeneous dimerized cubic lattice. The dependence on filling of the dimer singlet probability and of the entropy per site is shown. We set $U/t=5$ and $t_{\mathrm{d}}/t=10$ or $2$ and use different temperatures $k_{\mathrm{B}}T/t$. The entropy at half filling for large dimerization is strongly reduced.}
\label{sfig:2}
\end{figure*}

\begin{figure*}[ht!]
\includegraphics[scale=0.75]{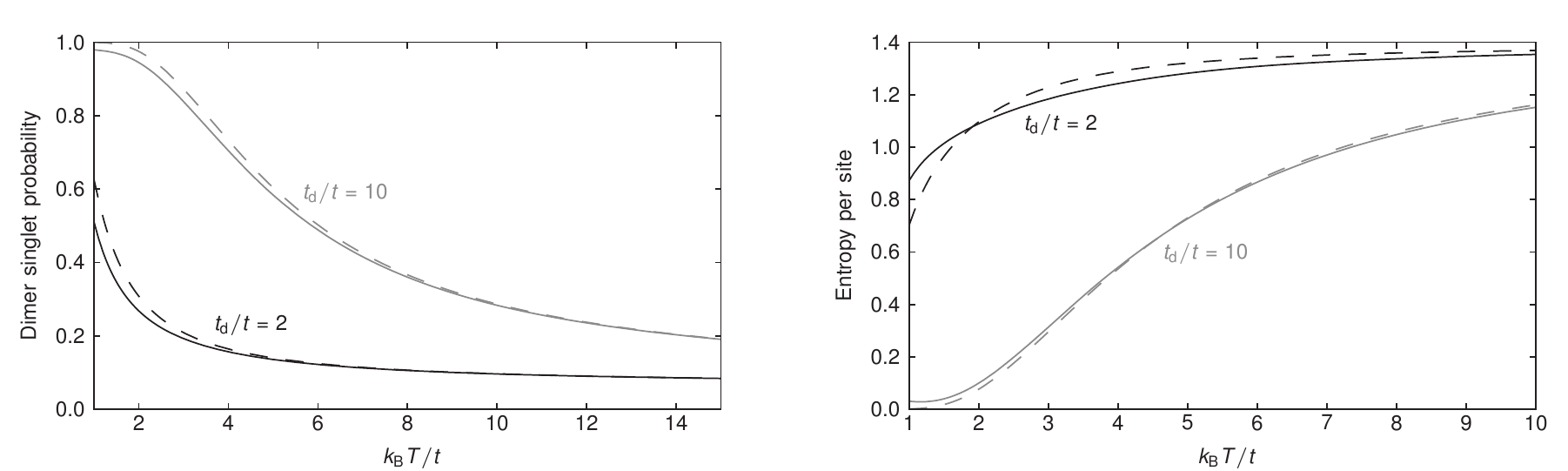}
\caption{
{\bf Higher order contributions.} Comparison of the high-temperature series predictions in lowest and second order (dashed and solid line) for the dimer singlet probability and for the entropy per site in a homogeneous dimerized cubic lattice. The filling is set to one particle per site and the interaction to $U/t=5$. The second order contributions are expected to be larger for lower temperatures and lead to a reduction of the singlet probability.}
\label{sfig:3}
\end{figure*}

As the expansion is up to second order in the tunnel coupling, it is sufficient to evaluate all expressions in a two-dimer basis. Denoting the single dimer Hamiltonian in the grand canonical potential with $\hat H_o^{\mathrm{s}}$, the unperturbed partition function then reads 
\begin{equation}
\mathrm{\cal{Z}}_0=\left(\mbox{Tr}\{e^{-\beta\hat H_{o}^{\mathrm{s}}}\}\right)^2.
\end{equation}
The evaluation of the second order terms is done in a double dimer basis $|\Psi^1_i,\Psi^2_j\rangle$, where $|\Psi^1_i\rangle$ and $|\Psi^2_j\rangle$ each denote one of the 16 possible eigenvectors of the first and second dimer link. This essentially leaves the evaluation of matrix elements of the following kind
\begin{eqnarray}
\langle\Psi^2_j,\Psi^1_i|\hat T'(\tau_1)\hat T'(\tau_2)|\Psi^1_i,\Psi^2_j\rangle & \qquad \mbox{and} \nonumber \\ \langle\Psi^2_j,\Psi^1_i|\mathrm{\cal{O}}\hat T'(\tau_1)\hat T'(\tau_2)|\Psi^1_i,\Psi^2_j\rangle,
\end{eqnarray}
which can be computed either directly or numerically. For the singlet and triplet fraction the observable $\mathrm{\cal{O}}$ takes the form of a projector for the 16 possible states on a dimer. The entropy and particle number per dimer are evaluated from the grand canonical potential $\Omega_{\mathrm{d}}=-k_{\mathrm{B}}T\mathrm{log}\mathrm{\cal{Z}}_d$ of a single dimer. Fig.~S2 shows the dimer singlet probability and the entropy per site versus filling calculated in second order for different temperatures and dimerizations. A comparison between the predictions of lowest order (atomic limit) and second order is shown in Fig. S3.

\section{Anisotropic cubic lattice}

Similar to the case of the dimerized lattice, we split the Hamiltonian for the homogeneous anisotropic cubic lattice into two parts
\begin{eqnarray}
\hat H_{A}&=&\hat H_{U}+\hat H_{t}\nonumber \\
    \hat H_{U}&=&U\sum_{i} \hat n_{i\uparrow}\hat n_{i\downarrow}-
    \mu\sum_{i} (\hat n_{i\uparrow}+\hat n_{i\downarrow}) \\
    \hat H_{t}&=&-t_s\sum_{\sigma,\left<i,j\right>_{\linethickness{2pt}\line(1,0){5}}} 
    (\hat c^\dagger_{i,\sigma}\hat c_{j,\sigma}+\text{h.c.}) 
    -t\sum_{\sigma,\left<i,j\right>_{\line(1,0){5}}} 
    (\hat c^\dagger_{i,\sigma}\hat c_{j,\sigma}+\text{h.c.}).\nonumber
\label{eq:fermi-hubbard2}
\end{eqnarray}
Notations are analogous to the previous section, see Fig.~S1. The strong tunneling between nearest neighbors along the $x$ direction is denoted with $t_s$, whereas the weaker tunneling along the other two axes is given by $t$. We treat the tunneling Hamiltonian $\hat H_t$ as a perturbation to the unperturbed part $\hat H_U$, which leads to an expansion of the partition function as in Eq. \ref{eq:hts} in powers of $\beta t_s$ and $\beta t$. Density and entropy per site are then obtained from derivatives of the second order grand potential $\Omega$,
\begin{equation}
\beta \Omega=-\mathrm{log}\mathrm{\cal{Z}}_1-\frac{2(\beta t_s)^2+4(\beta t)^2}{(\mathrm{\cal{Z}}_1)^2}\left(\zeta+\zeta^3w+2\zeta^2\frac{1-w}{\beta U}\right). 
\end{equation}
Here $\mathrm{\cal{Z}}_1$ is the unperturbed single site partition function, $\zeta=\mbox{exp}(\beta \mu)$ the fugacity and $w=\mbox{exp}(-\beta U)$. 

The evaluation of the two correlators defined in the main text $\langle S^z_iS^z_{i+1}\rangle$ and $\mathrm{\cal{S}}$ is slightly more complicated, as it involves two neighboring sites. However, the coefficients for these correlators have already been computed \cite{ten_haaf_high-temperature_1992}.
\begin{eqnarray}
\langle S^z_{i}S^z_{i+1}\rangle&=&-\frac{\zeta^2}{(\mathrm{\cal{Z}}_0^s)^2}\left(\frac{1}{\beta U}+\frac{w-1}{(\beta U)^2}\right)(\beta t_s)^2 \\
\mathrm{\cal{S}}&=&\left(\frac{1}{\beta U}+\frac{w-1}{(\beta U)^2}\right)(\beta t_s)^2.
\end{eqnarray}

\section{Harmonic trap}

The effect of the harmonic trap is included in a local density approximation with a quadratically varying chemical potential 
\begin{equation}
\mu(r)=\mu_0-\frac{1}{2}m\overline{\omega}^2(\frac{\lambda}{2})^2r^2, 
\end{equation}
where $\overline{\omega}$ is the geometric mean of the trapping frequencies, $\mu_0$ the chemical potential in the center of the trap and $r$ the normalized distance of a site to the trap center. Any trap averaged observable $\mathrm{\cal{O}}^{\mathrm{trap}}$ is then obtained from integration of the contributions per site $\mathrm{\cal{O}}^{\mathrm{hom}}(\mu)$
\begin{equation}
\mathrm{\cal{O}}^{\mathrm{trap}}=\int_0^\infty4\pi r^2\mathrm{\cal{O}}^{\mathrm{hom}}(\mu(r))\mbox{d}r.
\end{equation}
Owing to the harmonic trap, the energy offset between neighboring sites on the dimer links changes over the cloud size. The relative correction of this effect to all shown quantities was computed to be less than a few percent. 

\section{Singlet and triplet fractions}

\begin{figure*}[ht!]
\includegraphics[scale=0.75]{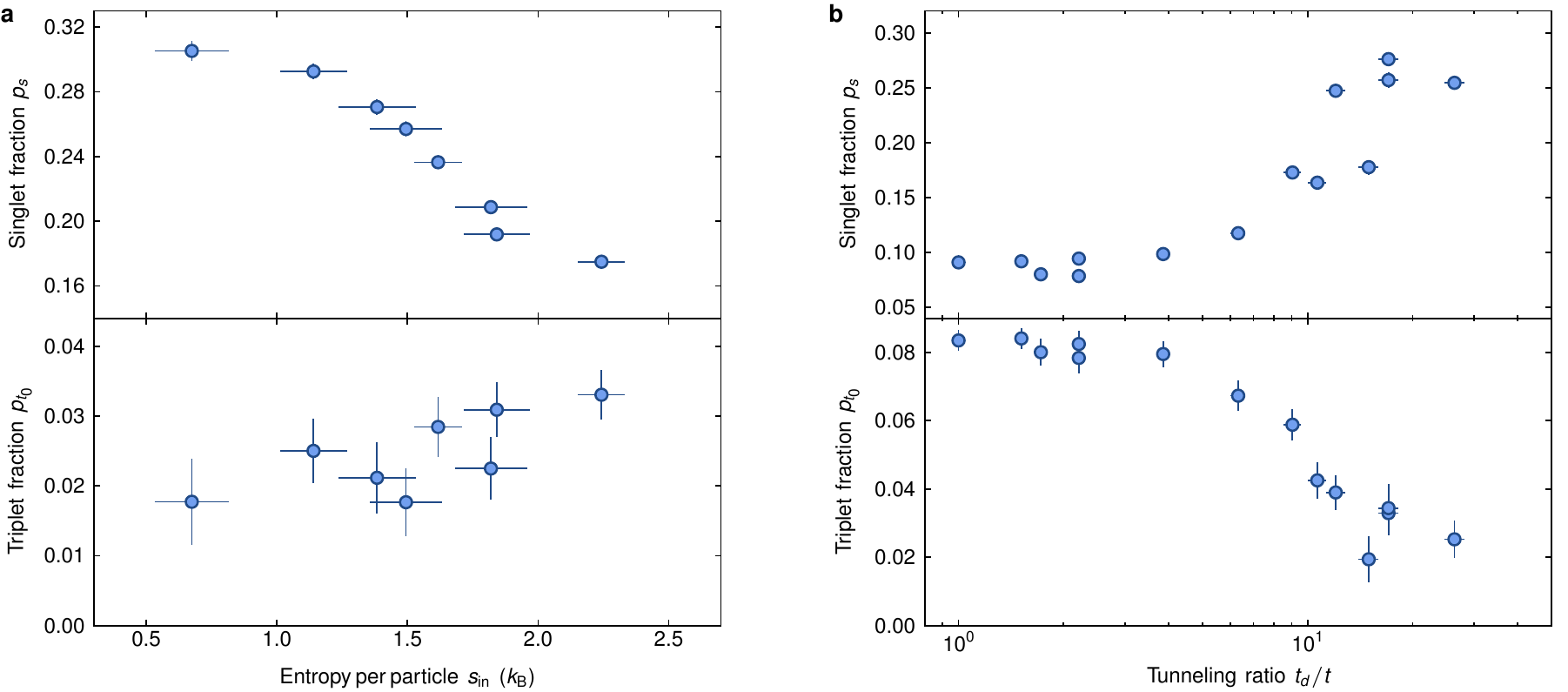}
\caption{
{\bf Dimerized simple cubic lattice.} 
{\bf a,} Singlet and triplet fractions on the strong dimer links vs. initial entropy before loading into the lattice. The data shown here is used to compute the normalized imbalance $\mathrm{\cal{I}}$ displayed in Fig.~3a.
{\bf b,} Singlet and triplet fractions as a function of dimerization $t_{\mathrm{d}}/t$, corresponding to the measurements of Fig.~3b.
Vertical error bars denote the fit error from singlet-triplet oscillations consisting of 63 measurements, the errors in $t_{\mathrm{d}}/t$ stem from lattice calibration uncertainties and the errors in $s_{\mathrm{in}}$ are the standard deviation of five measurements.}
\label{sfig:4}
\end{figure*}

\begin{figure*}[ht!]
\includegraphics[scale=0.75]{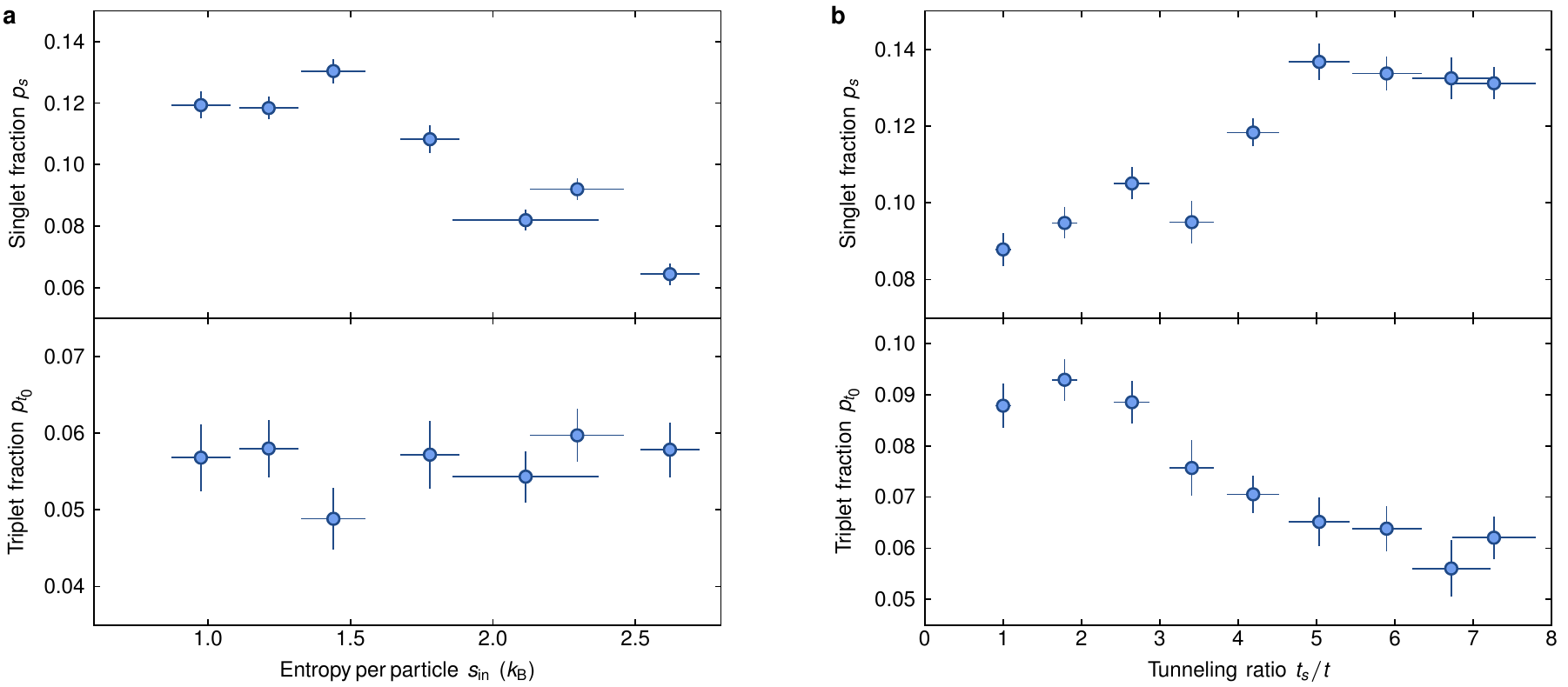}
\caption{
{\bf Anisotropic simple cubic lattice.} 
{\bf a,} Singlet and triplet fractions as a function of initial entropy before loading into the lattice, corresponding to the measurements of Fig.~4b.
{\bf b,} Singlet and triplet fractions vs. tunneling anisotropy $t_{\mathrm{s}}/t$, which are used to compute the spin correlators presented in Fig.~4a. 
Vertical error bars denote the fit error from singlet-triplet oscillations consisting of 63 measurements, the errors in $t_{\mathrm{s}}/t$ stem from lattice calibration uncertainties and the errors in $s_{\mathrm{in}}$ are the standard deviation of five measurements.}
\label{sfig:5}
\end{figure*}

The fraction of atoms forming singlets and triplets $p_{\mathrm{s}}$ and $p_{\mathrm{t_0}}$ are obtained from an integration over the left sites of each merged pair, which are part of the set $\mathcal{A}$ (see Fig.~2b),
\begin{alignat}{2}
p_{\mathrm{s}}&=2\sum_{i \in \mathcal{A}} \left<\hat P^{\mathrm{s}}_i\right>/N \qquad  &\hat P^{\mathrm{s}}_i=&\left|\Psi_i^{\mathrm{s}}\right>\left<\Psi_i^{\mathrm{s}}\right| \nonumber \\
p_{\mathrm{t_0}}&=2\sum_{i \in \mathcal{A}} \left<\hat P^{\mathrm{t_0}}_i\right>/N \qquad  &\hat P^{\mathrm{t_0}}_i=&\left|\Psi_i^{\mathrm{t_0}}\right>\left<\Psi_i^{\mathrm{t_0}}\right|.
\end{alignat}
Here $\hat P^{\mathrm{s}}_i$ and $\hat P^{\mathrm{t_0}}_i$ are the projection operators on the singlet and triplet states $\left|\Psi_i^{\mathrm{s}}\right>$ and $\left|\Psi_i^{\mathrm{t_0}}\right>$ on neighboring sites $i$ and $i+1$, $\left<...\right>$ denotes the thermal average and $N$ the total atom number. For the measurements in the anisotropic simple cubic lattice, the projection operators are related to the spin operators $\vec{S_i}=1/2\sum_{s,s'}\hat c^{\dagger}_{i,s}\vec{\sigma} c_{i,s'}$, where $\vec{\sigma}=(\sigma_x,\sigma_y,\sigma_z)$ are the Pauli matrices
\begin{eqnarray}
\hat P^{\mathrm{s}}_i &=&\frac{\hat n_i \hat n_{i+1}}{4}-\vec{S_i}\vec{S}_{i+1} \nonumber \\
\hat P^{\mathrm{t_0}}_i &=&\vec{S_i}\vec{S}_{i+1}-2S^z_iS^z_{i+1}+\frac{\hat n_i \hat n_{i+1}}{4}.
\end{eqnarray}
From this the equality of Eq.~(2) in the main text is immediately obtained. In the dimerized lattice the projection operator on the triplet reads the same, whereas for the singlet the two-site system needs to be diagonalized.

Fig.~S4 shows the individual atomic fractions of singlets and triplets ($p_{\mathrm{s}}$ and $p_{\mathrm{t_0}}$) measured in the dimerized lattice, which are used to compute the normalized imbalance $\mathrm{\cal{I}}$. The data in Fig.~S4a corresponds to the entropy scan of Fig.~3a in the main text, whilst the set of measurements for different tunneling ratios in Fig.~S4b corresponds to Fig.~3b. We additionally measured the fraction of atoms on doubly occupied sites $D_{\mathrm{dimer}}$ in the lattice after freezing out the atomic motion but before applying the cleaning procedure and inducing singlet-triplet oscillations. Note that for the dimer lattice, this fraction does not contain any contributions from the admixture of double occupancies present in the initial singlet state. For the entropy scan we find a nearly constant value of $D_{\mathrm{dimer}}=0.026(5)$. For the measurements taken at different $t_{\mathrm{d}}/t$ the double occupancy ranges between $0.066(4)$ and $0.29(2)$.

Fig.~S5 shows $p_s$ and $p_{t_0}$ for the measurements in the anisotropic cubic lattice, which are used to compute the spin correlators $-\langle S^{x}_{i}S^{x}_{i+1}\rangle-\langle S^{y}_{i}S^{y}_{i+1}\rangle$ and $\mathrm{\cal{S}}$. Fig.~S5a corresponds to the measurements versus entropy displayed in Fig.~4b, whilst the scan of the tunneling anisotropy shown in Fig.~S5b corresponds to the data presented in Fig.~4a. Here, we also measured the double occupancy immediately after freezing out the atomic motion $D_{\mathrm{anisotropic}}$, which now corresponds to a direct projection onto doubly occupied sites. For the entropy scan $D_{\mathrm{anisotropic}}$ lies between $0.11(6)$ and $0.21(1)$, whilst for the scan of $t_{\mathrm{s}}/t$ we measure double occupancies between $0.14(2)$ and $0.19(2)$.


\begin{thebibliography}{10}

\bibitem{auerbach_interacting_1994}
A.~Auerbach, {\it Interacting Electrons and Quantum Magnetism\/} (Springer,
  1994).

\bibitem{sachdev_quantum_2008}
S.~Sachdev, Quantum magnetism and criticality, {\it Nature Physics\/} {\bf 4},
  173 (2008).

\bibitem{anderson_physics_2004}
P.~W. Anderson, {\it et~al.\/}, The physics behind high-temperature
  superconducting cuprates: the plain vanilla version of {RVB}, {\it Journal of
  Physics: Condensed Matter\/} {\bf 16}, R755 (2004).

\bibitem{balents_spin_2010}
L.~Balents, Spin liquids in frustrated magnets, {\it Nature\/} {\bf 464}, 199
  (2010).

\bibitem{lewenstein_ultracold_2007}
M.~Lewenstein, {\it et~al.\/}, Ultracold atomic gases in optical lattices:
  mimicking condensed matter physics and beyond, {\it Advances in Physics\/}
  {\bf 56}, 243 (2007).

\bibitem{bloch_many-body_2008}
I.~Bloch, J.~Dalibard, W.~Zwerger, Many-body physics with ultracold gases, {\it
  Reviews of Modern Physics\/} {\bf 80}, 885 (2008).

\bibitem{esslinger_fermi-hubbard_2010}
T.~Esslinger, Fermi-Hubbard physics with atoms in an optical lattice, {\it
  Annual Review of Condensed Matter Physics\/} {\bf 1}, 129 (2010).

\bibitem{kohl_fermionic_2005}
M.~Köhl, H.~Moritz, T.~Stöferle, K.~Günter, T.~Esslinger, Fermionic atoms in
  a three dimensional optical lattice: observing Fermi surfaces, dynamics, and
  interactions, {\it Physical Review Letters\/} {\bf 94}, 080403 (2005).

\bibitem{jordens_mott_2008}
R.~Jördens, N.~Strohmaier, K.~Günter, H.~Moritz, T.~Esslinger, A Mott
  insulator of fermionic atoms in an optical lattice, {\it Nature\/} {\bf 455},
  204 (2008).

\bibitem{schneider_metallic_2008}
U.~Schneider, {\it et~al.\/}, Metallic and insulating phases of repulsively
  interacting fermions in a {3D} optical lattice, {\it Science\/} {\bf 322},
  1520 (2008).

\bibitem{simon_quantum_2011}
J.~Simon, {\it et~al.\/}, Quantum simulation of antiferromagnetic spin chains
  in an optical lattice, {\it Nature\/} {\bf 472}, 307 (2011).

\bibitem{struck_quantum_2011}
J.~Struck, {\it et~al.\/}, Quantum simulation of frustrated classical magnetism
  in triangular optical lattices, {\it Science\/} {\bf 333}, 996 (2011).

\bibitem{trotzky_time-resolved_2008}
S.~Trotzky, {\it et~al.\/}, Time-resolved observation and control of
  superexchange interactions with ultracold atoms in optical lattices, {\it
  Science\/} {\bf 319}, 295 (2008).

\bibitem{nascimbene_experimental_2012}
S.~Nascimbène, {\it et~al.\/}, Experimental realization of plaquette
  resonating valence-bond states with ultracold atoms in optical superlattices,
  {\it Physical Review Letters\/} {\bf 108}, 205301 (2012).

\bibitem{ho_squeezing_2009}
T.-L. Ho, Q.~Zhou, Squeezing out the entropy of fermions in optical lattices,
  {\it Proceedings of the National Academy of Sciences\/} {\bf 106}, 6916
  (2009).

\bibitem{bernier_cooling_2009}
J.-S. Bernier, {\it et~al.\/}, Cooling fermionic atoms in optical lattices by
  shaping the confinement, {\it Physical Review A\/} {\bf 79}, 061601 (2009).

\bibitem{diep_two-dimensional_2005}
H.~T. Diep, G.~Misguich, C.~Lhuillier, {\it Frustrated Spin Systems\/} (World
  Scientific, 2005), pp. 229--306.

\bibitem{tarruell_creating_2012}
L.~Tarruell, D.~Greif, T.~Uehlinger, G.~Jotzu, T.~Esslinger, Creating, moving
  and merging Dirac points with a Fermi gas in a tunable honeycomb lattice,
  {\it Nature\/} {\bf 483}, 302 (2012).

\bibitem{supplementary}
{S}ee {S}upplementary {M}aterials.

\bibitem{trotzky_controlling_2010}
S.~Trotzky, Y.-A. Chen, U.~Schnorrberger, P.~Cheinet, I.~Bloch, Controlling and
  detecting spin correlations of ultracold atoms in optical lattices, {\it
  Physical Review Letters\/} {\bf 105}, 265303 (2010).

\bibitem{fuchs_thermodynamics_2011}
S.~Fuchs, {\it et~al.\/}, Thermodynamics of the {3D} Hubbard model on
  approaching the Néel transition, {\it Physical Review Letters\/} {\bf 106},
  030401 (2011).

\bibitem{greif_probing_2011}
D.~Greif, L.~Tarruell, T.~Uehlinger, R.~Jördens, T.~Esslinger, Probing
  nearest-neighbor correlations of ultracold fermions in an optical lattice,
  {\it Physical Review Letters\/} {\bf 106}, 145302 (2011).

\bibitem{jordens_quantitative_2010}
R.~Jördens, {\it et~al.\/}, Quantitative determination of temperature in the
  approach to magnetic order of ultracold fermions in an optical lattice, {\it
  Physical Review Letters\/} {\bf 104}, 180401 (2010).

\bibitem{ruegg_pressure-induced_2004}
C.~Rüegg, {\it et~al.\/}, Pressure-induced quantum phase transition in the
  spin-liquid {$\mathrm{TlCuCl}_{3}$}, {\it Physical Review Letters\/} {\bf
  93}, 257201 (2004).

\bibitem{werner_interaction-induced_2005}
F.~Werner, O.~Parcollet, A.~Georges, S.~R. Hassan, Interaction-induced
  adiabatic cooling and antiferromagnetism of cold fermions in optical
  lattices, {\it Physical Review Letters\/} {\bf 95}, 056401 (2005).

\bibitem{mathy_accessing_2009}
C.~J.~M. Mathy, D.~A. Huse, Accessing the Néel phase of ultracold fermionic
  atoms in a simple-cubic optical lattice, {\it Physical Review A\/} {\bf 79},
  063412 (2009).

\bibitem{ma_density_2012}
P.~N. Ma, S.~Pilati, M.~Troyer, X.~Dai, Density functional theory for atomic
  Fermi gases, {\it Nature Physics\/} {\bf 8}, 601 (2012).

\bibitem{giamarchi_quantum_2003}
T.~Giamarchi, {\it Quantum Physics in One Dimension\/} (Clarendon Press, 2003).

\bibitem{gorelik_universal_2012}
E.~V. Gorelik, {\it et~al.\/}, Universal probes for antiferromagnetic
  correlations and entropy in cold fermions on optical lattices, {\it Physical
  Review A\/} {\bf 85}, 061602 (2012).

\bibitem{he_magnetic_2007}
P.-B. He, Q.~Sun, P.~Li, S.-Q. Shen, W.~M. Liu, Magnetic quantum phase
  transition of cold atoms in an optical lattice, {\it Physical Review A\/}
  {\bf 76}, 043618 (2007).

\bibitem{meng_quantum_2010}
Z.~Y. Meng, T.~C. Lang, S.~Wessel, F.~F. Assaad, A.~Muramatsu, Quantum spin
  liquid emerging in two-dimensional correlated Dirac fermions, {\it Nature\/}
  {\bf 464}, 847 (2010).

\bibitem{krauser_coherent_2012}
J.~S. Krauser, {\it et~al.\/}, Coherent multi-flavour spin dynamics in a
  fermionic quantum gas, {\it Nature Physics\/} {\bf 8}, 813 (2012).

\bibitem{henderson_high-temperature_1992}
J.~A. Henderson, J.~Oitmaa, M.~C.~B. Ashley, High-temperature expansion for the
  single-band Hubbard model, {\it Physical Review B\/} {\bf 46}, 6328 (1992).

\bibitem{ten_haaf_high-temperature_1992}
D.~F.~B. ten Haaf, J.~M.~J. van Leeuwen, High-temperature series expansions for
  the Hubbard model, {\it Physical Review B\/} {\bf 46}, 6313 (1992).

\end{thebibliography}
\end{document}